\documentclass[aps,prl,twocolumn,superscriptaddress,longbibliography]{revtex4-2}
\usepackage{graphicx}
\usepackage{dcolumn}   
\usepackage{bm}        
\usepackage{amssymb}   
\usepackage{amsmath}
\usepackage{url}
\usepackage{layout}
\usepackage{color}
\usepackage{epsfig}
\usepackage{graphicx}
\usepackage{cancel}
\usepackage{mathrsfs}\usepackage{bm}\usepackage{appendix}\usepackage{color}

\usepackage[thinlines]{easytable}
\usepackage{makecell}
\usepackage{tikz,pgf}
\usepackage{booktabs}
\usepackage{float}
\usepackage{subfigure}
\usepackage{hyperref}
\usepackage{lineno}
\hypersetup{colorlinks,allcolors=blue}

\begin{document}

\title{Band Structure and Pairing Nature of La$_3$Ni$_2$O$_7$ Thin Film at Ambient Pressure}
\author{Zhi-Yan Shao}
\thanks{These two authors contributed equally to this work.}
\affiliation{School of Physics, Beijing Institute of Technology, Beijing 100081, China}
\author{Yu-Bo Liu}
\thanks{These two authors contributed equally to this work.}
\affiliation{Institute of Theoretical Physics, Chinese Academy of Sciences, Beijing 100190, China}
\author{Min Liu}
\email{minliu@buct.edu.cn}
\affiliation{College of Mathematics and Physics, Beijing University of Chemical Technology, Beijing 100029, China}
\author{Fan Yang}
\email{yangfan\_blg@bit.edu.cn}
\affiliation{School of Physics, Beijing Institute of Technology, Beijing 100081, China}

\begin{abstract}
Recently, evidences of superconductivity (SC) with onset $T_c$ above the McMillan limit have been detected in the La$_3$Ni$_2$O$_7$ ultrathin film grown on the LaSrAlO$_4$ substrate at ambient pressure. This progress opens a new era in the field of the nickelate superconductors. Here we perform a density-functional-theory (DFT) based calculation for the band structure of this material. The obtained DFT+$U$ band structure has the feature that the bonding $d_{z^2}$ band crosses the Fermi level, forming the hole pocket $\gamma$, consistent with the angle-resolved photoemission spectrum (ARPES). Taking the low-energy Ni-$(3d_{z^2},3d_{x^2-y^2})$ orbitals placed on the tetragonal lattice structure, we construct a 2D bilayer four-band tight-binding model which well captures the main features of the DFT+$U$ band structure. Then considering the multi-orbital Hubbard interaction, we adopt the random-phase approximation (RPA) approach to investigate the pairing nature. The obtained pairing symmetry is $s^{\pm}$ or $d_{xy}$ for the hole-doping level $\delta$ below or above 0.12, induced by the different Fermi surface nesting situations. For the realistic $\delta=0.21$ measured by the ARPES, our RPA calculations obtain the next-nearest-neighbor pairing $d_{xy}$-wave SC dominated by the $d_{z^2}$ orbital, consistent with the experimental observation that the $T_c$ enhances with the shrinking of the in-plane lattice constants. This pairing state is induced by the nesting between the different patches within the $\gamma$ pocket. Our results appeal for experimental verifications.
\end{abstract}

\maketitle

\paragraph{{\color{blue}Introduction:}} The discovery of high-temperature superconductivity (SC) in the Ruddlesden-Popper (RP) phase multilayer nickelate superconductors La$_3$Ni$_2$O$_7$~\cite{Wang2023LNO} and La$_4$Ni$_3$O$_{10}$~\cite{zhu2024superconductivity,zhang2023superconductivity,huang2024signature,li2023trilayer} under high pressure (HP) has aroused a surge in the exploration of the pairing mechanism and physical properties of the nickelates family both experimentally~\cite{YuanHQ2023LNO,Wang2023LNOb,wang2023LNOpoly,wang2023la2prnio7,zhang2023pressure,zhou2023evidence,wang2024bulk,li2024pressure,puphal2024unconven,Dong2024vis,Chen2024poly,wang2024chemical,Li2024ele,li2024distinguishing,zhou2024revealing,cui2023strain,yang2024orbital,wang2023structure,huo2025modulation} and theoretically~\cite{cao2023flat,WangQH2023,YangF2023,lu2023bilayertJ,oh2023type2,qu2023bilayer,Yi_Feng2023,jiang2023high,qin2023high,zhang2023strong,pan2023rno,yang2023strong,fan2023sc,wu2024deconfined,zhang2024prediction,zhang2024s,Yang2024effective,zhang2024electronic,yang2024decom,Lu2024interplay,ZhangGM2023DMRG,heier2023competing,
sui2023rno,YaoDX2023,Dagotto2023,cao2023flat,zhang2023structural,huang2023impurity,geisler2023structural,rhodes2023structural,zhang2023la3ni2o6,yuan2023trilayer,li2024la3,geisler2024optical,li2017fermiology,wang2024non,chen2024tri,Li2024design,WangQH2023,YangF2023,lechermann2023,Kuroki2023,HuJP2023,lu2023bilayertJ,oh2023type2,liao2023electron,qu2023bilayer,Yi_Feng2023,jiang2023high,zhang2023trends,qin2023high,tian2023correlation,jiang2023pressure,lu2023sc,kitamine2023,luo2023high,zhang2023strong,pan2023rno,sakakibara2023La4Ni3O10,lange2023mixedtj,yang2023strong,lange2023feshbach,kaneko2023pair,fan2023sc,wu2024deconfined,zhang2024prediction,zhang2024s,Yang2024effective,zhang2024electronic,yang2024decom,ryee2024quenched,Lu2024interplay,Ouyang2024absence,ZhangGM2023DMRG,Werner2023,shilenko2023correlated,WuWei2023charge,chen2025charge,ouyang2023hund,heier2023competing,wang2024electronic,botzel2024theory,gao2025robust}. As a new platform to study the high-temperature SC, the nickelates family is different from the cuprates and the iron-based superconductors family in their multilayer structure, which is believed to play a crucial role in the pairing mechanism~\cite{cao2023flat,WangQH2023,YangF2023,lu2023bilayertJ,oh2023type2,qu2023bilayer,Yi_Feng2023,jiang2023high,qin2023high,zhang2023strong,pan2023rno,yang2023strong,fan2023sc,wu2024deconfined,zhang2024prediction,zhang2024s,Yang2024effective,zhang2024electronic,yang2024decom,Lu2024interplay,ZhangGM2023DMRG,heier2023competing}. Previously, the SC in these RP phase nickelates only emerges under HP, while most experiments can only be conducted at ambient pressure (AP) due to technical difficulties. This strongly hinders the experimental investigation of the pairing mechanism of these superconductors. Very recently, this research field witnessed a breakthrough, i.e. the detection of SC with onset $T_c$ above the McMillan limit ($\approx$40 K) in the La$_3$Ni$_2$O$_7$~\cite{Ko2024signature}, La$_{2.85}$Pr$_{0.15}$Ni$_2$O$_7$~\cite{zhou2024ambient} and La$_2$PrNi$_2$O$_7$ \cite{liu2025superconductivity} ultrathin films grown on the LaSrAlO$_4$ (LSAO) substrate at AP by two different teams.  

While the zero resistivity and Meissner effect detected by both teams identify the presence of real SC, the Brezinskii-Kosterlitz-Thouless transition character and the anisotropic critical magnetic fields reveal the 2D characteristic of the SC in these bilayer nickelate films~\cite{Ko2024signature,zhou2024ambient,liu2025superconductivity}. Experimental tools such as the scanning transmission electron microscopy (STEM) have revealed that these film materials host the same tetragonal crystal structure as that of the bulk materials under HP~\cite{bhatt2025resolving,yue2025correlated}, in which there is no tilting of the NiO$_6$ octahedra. Through tuning the strength of the strain, it is found that the $T_c$ enhances with the shrinking of in-plane lattice constant and is insensitive to the c-axis one~\cite{Ko2024signature,bhatt2025resolving}. In the aspect of electronic structure, the angle-resolved photoemission spectrum (ARPES) identifies a band structure qualitatively consistent with the density-functional-theory (DFT) $+U$ calculations~\cite{yue2025correlated,li2025photoemission}. Notably, the bonding $d_{z^2}$ band top crosses the Fermi level, forming a hole pocket $\gamma$ centering around the $M(\pi,\pi)$-point similar with the case in the bulk La$_3$Ni$_2$O$_7$ under HP. Significantly, from the area enclosed by the Fermi surfaces (FSs) measured by ARPES, it is found that hole-doping has been introduced to the film, probably through Sr diffusion from the substrate~\cite{li2025photoemission}. Currently, the pairing nature of this material remains unknown.  

In this paper, we construct a Ni-$(3d_{z^2},3d_{x^2-y^2})$ orbital bilayer tight-binding (TB) model to investigate the pairing nature of the La$_3$Ni$_2$O$_7$ ultrathin film. Our DFT$+U$ band structure shows that the bonding-$d_{z^2}$ band crosses the Fermi level, forming the $\gamma$-pocket, consistent with the ARPES observation. Using the two low-energy Ni-$3d$-$\mathrm{e_g}$ orbitals placed in the tetragonal structure, we construct a 2D four-band TB model, which well fits the DFT band structure. After considering the multi-orbital Hubbard interactions, we engage the random-phase-approximation (RPA) approach to investigate the pairing symmetry. Our RPA results suggest that the pairing symmetry is $s^{\pm}$ or $d_{xy}$ for the hole-doping level $\delta$ below or above 0.12. For the realistic $\delta=0.21$ measured by ARPES, the $d_{xy}$-wave pairing is the leading pairing symmetry, which is induced by the nesting within the $\gamma$-pocket. The real-space pairing pattern is dominated by the next-nearest-neighbor (NNN) pairing between the $d_{z^2}$ orbitals.  Our results appeal for experimental verifications.

\begin{figure}[htbp]
		\centering
		\includegraphics[width=0.45\textwidth]{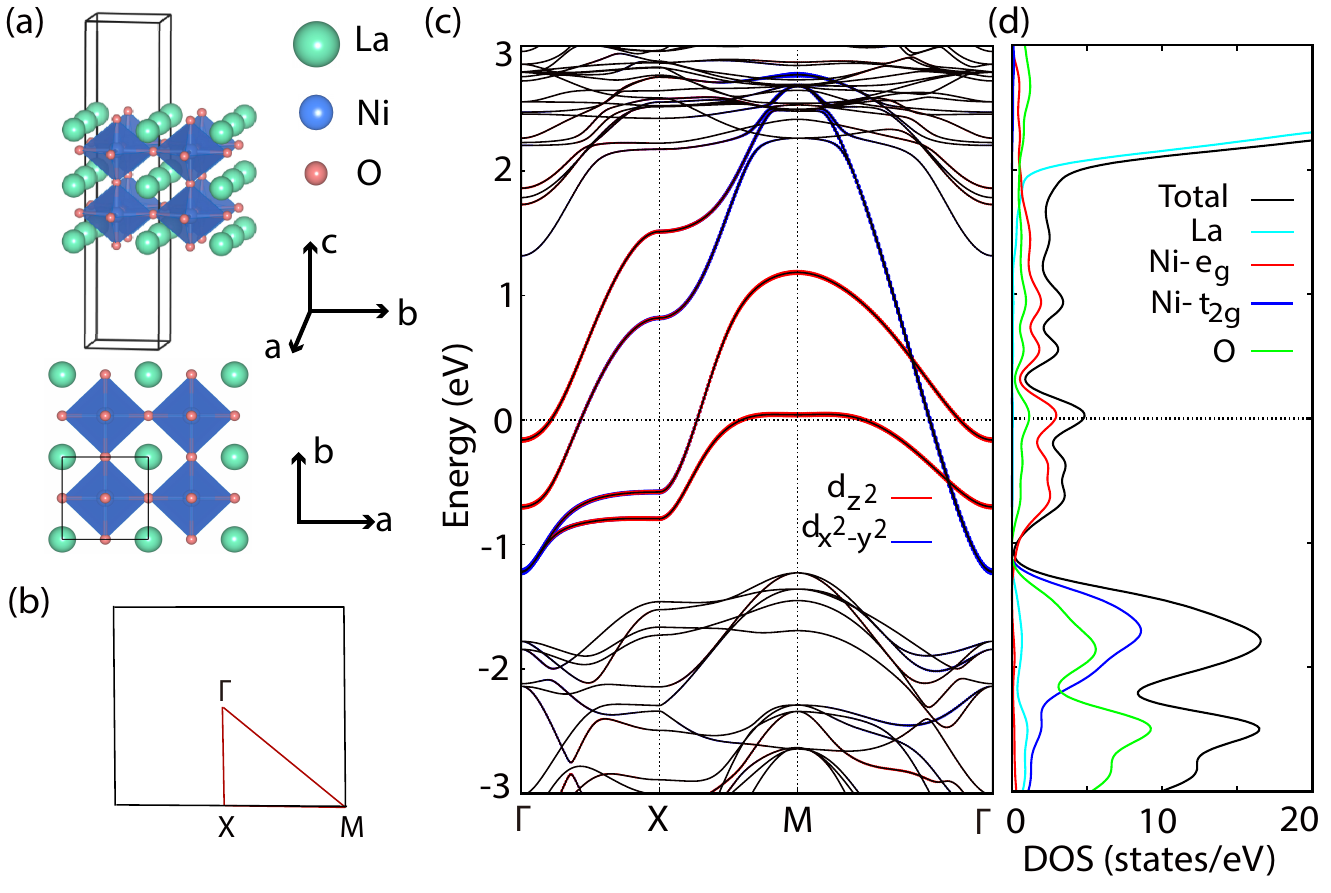}
		\caption{(color online) The DFT electronic structures. (a) Side- (upper) and top- (lower) views of the crystal structure of the La$_3$Ni$_2$O$_7$ single bilayer. (b) 2D BZ with high-symmetric points and lines marked. (c) DFT+$U$ band structure of the La$_3$Ni$_2$O$_7$ single-bilayer with $U$ = 3.5 eV. The blue and red colors represent for the weights of Ni-$d_{x^2-y^2}$ and Ni-$d_{z^2}$ orbitals respectively. (d) The DOS contributed by La, O and Ni-3$d$ orbitals. }
		\label{fig1}
\end{figure}


\paragraph{{\color{blue}DFT Band Structure and TB Model:}} As seen in Fig.~\ref{fig1}(a), the ultrathin film of La$_3$Ni$_2$O$_7$ crystallizes in a tetragonal phase at AP, in which there is no tilting of the NiO$_6$ octahedron. We consider a single-bilayer. First-principles DFT calculations were implemented utilizing the Vienna ab initio simulation package (VASP) and employing the projector augmented wave (PAW) as the pseudopotentials~\cite{kress1993vasp,kress1996planewave,blochl1994paw}. The electronic correlations were considered by the generalized gradient approximation (GGA) and the Perdew-Burke-Ernzerhof (PBE) exchange potential~\cite{perdew1996gga}. The plane-wave cutoff energy was set as 550 eV and a $k$-point grid 12 × 12 × 1 was adopted for the single-bilayer structure of La$_3$Ni$_2$O$_7$ with the $p4/mm$ symmetry reported in Ref. \cite{yue2025correlated}, where $a=b=3.7544\ \mathrm{\AA}$ and the spacing between the two NiO$_2$ layers is $4.28\ \mathrm{\AA}$. We performed DFT+$U$ band calculations with a reasonable onsite Coulomb interaction $U$ = 3.5 eV~\cite{Wang2023LNO,yang2024orbital} to correct the band structure in strongly correlated electron systems. 

Along the high-symmetric lines exhibited in the 2D Brillouin zone (BZ) shown in Fig.~\ref{fig1}(b), we present our DFT+$U$ band structure in Fig.~\ref{fig1}(c). The orbital weight distribution shown in Fig.~\ref{fig1}(c) suggests that the electronic states near the Fermi level are primarily derived from the two Ni-$3d$-e$_\text{g}$ orbitals, i.e. $d_{x^2-y^2}$ and $d_{z^2}$, as is also verified in the orbital-dependent density of state (DOS) shown in Fig.~\ref{fig1}(d). Owing to the interlayer coupling through the Ni-$d_{z^2}$ and O-$p_z$ orbitals, the Ni-$d_{z^2}$ bands split into the bonding and antibonding states around the $M$ point. However, the interlayer coupling is reduced here due to the elongated NiO$_2$ bilayer distance~\cite{Ko2024signature,zhou2024ambient,yue2025correlated}, leading to the reduced band splitting. Consequently, the bonding $d_{z^2}$ band crosses the Fermi level, forming a hole pocket, consistent with the ARPES observation~\cite{li2025photoemission}. This feature, in comparison with bulk La\textsubscript{3}Ni\textsubscript{2}O\textsubscript{7}, is different from the case at AP~\cite{liu2024origin}, but similar with the case under HP~\cite{YaoDX2023}.

\begin{figure}[htbp]
		\centering
		\includegraphics[width=0.4\textwidth]{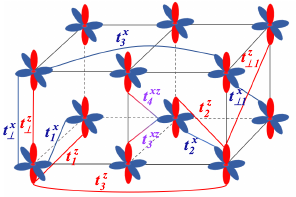}
		\caption{(color online) Schematic of the hopping integrals for La$_3$Ni$_2$O$_7$ film at AP.}
		\label{ambient}
\end{figure}

\begin{figure}[htbp]
		\centering
		\includegraphics[width=0.45\textwidth]{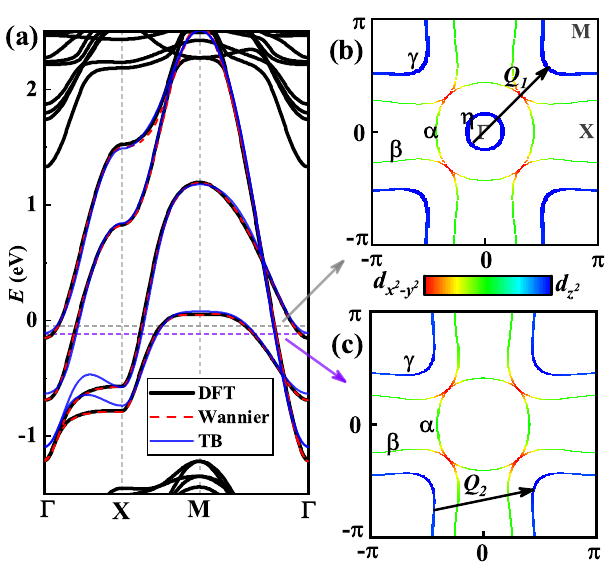}
		\caption{(color online) 
Band structure and FSs of La$_3$Ni$_2$O$_7$ film at AP obtained from the four-band TB-model Eq.~(\ref{equTB}). (a) The band structure of model Eq.~(\ref{equTB}) (blue lines) along the high symmetry lines, compared with the DFT band (black lines) and Wannier band (red dashed lines). The gray and purple dashed lines represent the Fermi levels for the hole-doping level $\delta=0.1$ and $0.23$, respectively.
(b-c) FSs in the BZ with hole-doping levels $\delta=0.1$ (b) and $\delta=0.23$ (c). In (b-c), The FSs marked as $\alpha$, $\beta$, $\gamma$ and $\eta$. The color indicates the orbital weight of $d_{x^2-y^2}$ and $d_{z^2}$. The FS-nesting vector is marked by $\bm{Q_1}/\bm{Q_2}$ in (b/c).}
		\label{tb}
\end{figure}

For convenience in the subsequent studies, we construct a Wannier model with the Ni-$(d_{x^2-y^2}, d_{z^2})$ orbitals based on the maximally-localized Wannier function method implemented in the WANNIER90 code~\cite{mostofi2008wannier90}. The hopping parameters in the TB model are obtained by Wannier downfolding the band structure with Ni-$d_{x^2-y^2}$ and Ni-$d_{z^2}$ orbitals. As shown in  Fig.~\ref{tb}(a), the Wannier bands align remarkably well with the DFT bands near the Fermi level. 

\begin{table}[htbp]
 \renewcommand{\arraystretch}{1.5}
 \setlength{\tabcolsep}{5pt}
\centering
\caption{The hopping integrals of the $(d_{z^2}, d_{x^2-y^2})$-orbital bilayer TB model for the La$_3$Ni$_2$O$_{7}$ film at AP. In the superscript and subscript, $x(z)$ represents the $d_{x^2-y^2}(d_{z^2})$ orbit, $\perp$ represents interlayer hopping, and $1,2,3$ represents the nearest-neighbor, next-nearest-neighbor and third-nearest-neighbor hopping, respectively. $\epsilon$ is on-site energy. The unit of all parameters is eV.}
\begin{tabular}{c|c|c|c|c}
    \hline\hline
    $t_{1}^{x}$ & $t_{2}^{x}$ & $t_{3}^{x}$ & $t_{\perp}^{x}$ & $t_{\perp}^{x1}$ \\
    \hline
    -0.4501 & 0.0615 & -0.0651 & 0.0027 & 0 \\
    \hline
    $t_{1}^{z}$ & $t_{2}^{z}$ & $t_{3}^{z}$ & $t_{\perp}^{z}$ & $t_{\perp}^{z1}$ \\
    \hline
    -0.1248 & -0.0205 & -0.0110 & -0.4060 & 0.0360 \\
    \hline
    $t_{3}^{xz}$ & $t_{4}^{xz}$ & $\epsilon_{x}$ & $\epsilon_{z}$ & \\
    \hline
    0.2217 & -0.0289 & 0.8898 & 0.4218 & \\
    \hline\hline
\end{tabular}
\label{tab2}
\end{table}

\paragraph{{\color{blue}TB Model and Microscopic Hamiltonian:}}Considering the hoppings up to a few neighborings, the TB model can be obtained and transformed into the $\bm{k}$-space,
\begin{equation}
    \begin{aligned}
        H_{\mathrm{TB}} = & \sum_{ij\mu\nu\alpha\beta\sigma} t_{i\mu\alpha,j\nu\beta} c^{\dag}_{i\mu\alpha\sigma}c_{j\nu\beta\sigma} \\
        = & \sum_{\bm{k}\mu\nu\alpha\beta\sigma} \left(H_{\bm{k}}\right)_{\mu\alpha,\nu\beta} c^{\dag}_{\bm{k}\mu\alpha\sigma}c_{\bm{k}\nu\beta\sigma}.
    \end{aligned}\label{equTB}
\end{equation}
Here $i/j$ labels site, $\mu/\nu$ labels layer, $\alpha/\beta$ labels orbital ($d_{z^2}$ ($z$) and $d_{x^2-y^2}$ ($x$)), $\sigma$ labels the spin, the $t_{i\mu\alpha,j\nu\beta}$ represent for the hopping integrals, whose definitions are illustrated in Fig. \ref{ambient} and the corresponding values are listed in Tab. \ref{tab2}. $H_{\bm{k}}$ is the $\bm{k}$-space TB Hamiltonian matrix. The effective orbital indices is the combined layer ($t$/$b$) and physical orbital indices ($x$/$z$).
In the basis $(tz, tx, bz, bx)$, the $k$-space TB Hamiltonian $H_{\bm{k}}$ takes the form
\begin{equation}
    H_{\bm{k}} = \left[
    \begin{array}{cc}
        H_1 & H_2 \\
        H_2^{\dag} & H_1
    \end{array}
    \right].
\end{equation}
Here the blocks $H_1$ and $H_2$ are expressed as
\begin{equation}
    H_1 = \left(
    \begin{array}{cc}
        h_{1}^{z} & h_2 \\
        h_2 & h_1^{x}
    \end{array}
        \right) ,\ 
    H_2 = \left(
    \begin{array}{cc}
        h_3^z & h_4 \\
        h_4 & h_3^x
    \end{array}
    \right),
\end{equation}
where
\begin{align}
    h_{1}^{\alpha} & = 2 t_1^{\alpha} (\cos{}k_x+\cos{}k_y) + 2 t_3^{\alpha} [\cos(2k_x)+\cos(2k_y)] \nonumber\\ & + 2 t_2^{\alpha} [\cos(k_x+k_y)+\cos(k_x-k_y)] + \epsilon_{\alpha} \ (\alpha=z,x), \nonumber\\
    h_2 & = 2 t_3^{xz} (\cos{}k_x-\cos{}k_y), \nonumber\\
    h_3^{\alpha} & = t_{\perp}^{\alpha} + 2 t_{\perp1}^{\alpha} (\cos{}k_x+\cos{}k_y) \ (\alpha=z,x), \nonumber\\
    h_4 & = 2t_4^{xz} (\cos{}k_x-\cos{}k_y),
\end{align}
where $k_{x}$ and $k_{y}$ are the $x$- and $y$- components of $\bm{k}$.

In comparison with the TB parameters for bulk La\textsubscript{3}Ni\textsubscript{2}O\textsubscript{7} at AP~\cite{huo2025modulation}, the intralayer nearest-neighbor hopping of the $d_{x^2-y^2}$ electrons $t_{1}^{x}$ becomes stronger due to the shrinked in-plane lattice constants, while the interlayer hopping of the $d_{z^2}$ electrons $t_{\perp}^{z}$ gets weaker due to the elongated c-axis one. 

The band structure of our four-band TB model Eq.~(\ref{equTB}) is shown in Fig. \ref{tb}(a), in comparison with the DFT and Wannier ones.  Clearly, the former has captured the main features of the latter. As the ARPES result suggests that the system is hole-doped~\cite{li2025photoemission}, we show the FSs of two typical hole-doping levels $\delta=0.1$ and $0.23$ in Fig. \ref{tb} (b) and (c), which correspond to the average particle number per site $n=1.5-\delta=1.4$ and $1.3$, respectively. In both FSs, there exist electron pocket $\alpha$ centering around the $\Gamma$ point and hole pockets $\beta$ and $\gamma$ centering around the M point. In the $\delta=0.1$ case, a small electron pocket $\eta$ centering around the $\Gamma$ point also exists. The $\alpha$ and $\beta$ pockets show mixing of orbital contents while the $\gamma$ and $\eta$ pockets are dominated by the $d_{z^2}$ orbital. Although the FS topologies of the two dopings are the same, their FS-nesting vectors are distinct due to their different pocket sizes. For $\delta=0.1$ the dominant nesting is between the $\eta$ and $\gamma$ pockets with nesting vector $\bm{Q}_1\approx(0.67\pi,0.67\pi)$; while for $\delta=0.23$ the dominant nesting is between the different parts of the $\gamma$ pocket with nesting vector $\bm{Q}_2\approx(0.9\pi,0.2\pi)$.

\begin{figure}[htbp]
		\centering
		\includegraphics[width=0.47\textwidth]{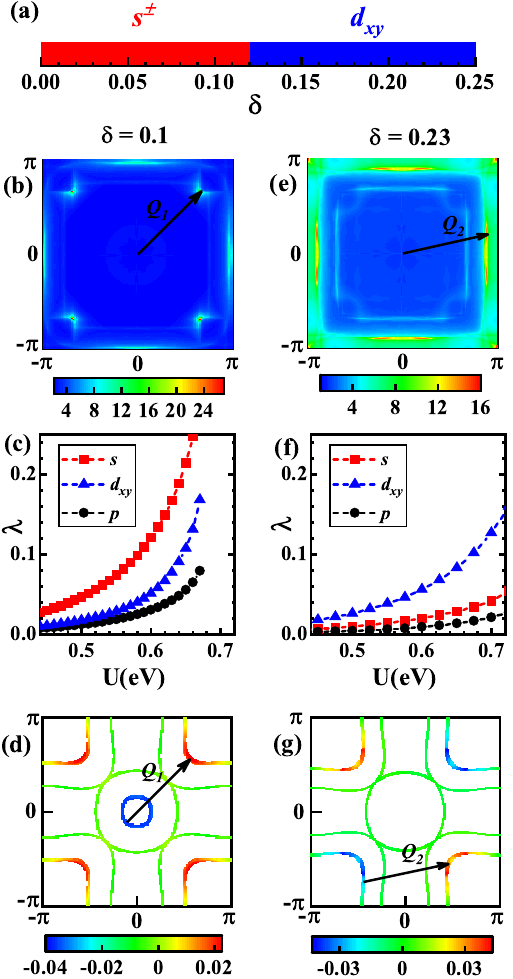}
		\caption{(color online) (a) The leading pairing symmetry as function of the hole-doping level $\delta=1.5-n$. (b-d) The RPA results for $\delta=0.1$. (b) The distribution of the spin susceptibility $\chi^{(s)}(\bm{k})$ over the BZ for $U<U_c$. $\bm{Q_1}\approx (0.67\pi,0.67\pi)$ is just the FS-nesting vector for this doping. (c)  The largest pairing eigenvalue $\lambda$ as function of $U$ for different pairing symmetries. (d) Distribution of the leading $s$-wave pairing gap function on the FSs. (e-g) The RPA results for $\delta=0.23$. (e) The distribution of the spin susceptibility $\chi^{(s)}(\bm{k})$ over the BZ for $U<U_c$. $\bm{Q_2}\approx (0.9\pi,0.2\pi)$ is the FS-nesting vector for this doping. (f)  The largest pairing eigenvalue $\lambda$ as function of $U$ for different pairing symmetries. (g) Distribution of the leading $d_{xy}$-wave pairing gap function on the FSs.}
		\label{rpa}
\end{figure}

\paragraph{{\color{blue}Doping-Dependent Pairing Symmetries:}} After considering the electron-electron interaction, we obtain the following multi-orbital Hubbard model,
\begin{equation} \label{eq_Hubbard}
    \begin{aligned}
       H=& H_{\text{TB}}+U \sum_{i\mu\alpha} n_{i\mu\alpha\uparrow}n_{i\mu\alpha\downarrow} + V \sum_{i\mu} n_{i\mu{}z}n_{i\mu{}x} \\ & + J_H \sum_{i\mu} \left[ \left( \sum_{\sigma\sigma^{\prime}} c^{\dag}_{i\mu{}z\sigma}c^{\dag}_{i\mu{}x\sigma^{\prime}}c_{i\mu{}z\sigma^{\prime}}c_{i\mu{}x\sigma} \right) \right. \\ & \phantom{+J_H\sum_{i\mu}} \left. + \left( c^{\dag}_{i\mu{}z\uparrow}c^{\dag}_{i\mu{}z\downarrow}c_{i\mu{}x\downarrow}c_{i\mu{}x\uparrow} + \mathrm{h.c.} \right) \right].
    \end{aligned}
\end{equation}
Here the $U$ and $V$ terms are the intraorbital and interorbital repulsions, and the $J_H$ term includes the Hund's coupling and pair hopping. 
Note that $U$ here is a parameter representing the residual Coulomb interaction between quasiparticles after screening, whose value can be different from the one in the DFT$+U$ study. 
We use the relation $U=V+2J_H$ and set $J_H=U/6$ in our study. This Hamiltonian is solved by standard multi-orbital RPA approach~ \cite{takimoto2004strong,yada2005origin,kubo2007pairing,graser2009near,liu2013d+,zhang2022lifshitz,kuroki101unconventional} (see Appendix A for details). In the RPA framework, the SC is driven by the spin fluctuation whose propagator is given by the spin susceptibility renormalized up to the RPA level. 
When $U$ is below the critical interaction strength $U_c$, the $T_c$ of the SC mediated by the spin fluctuation is determined by the pairing eigenvalue $\lambda$ through $T_c\sim e^{-1/\lambda}$, and the pairing symmetry is determined by the corresponding pairing eigenvector. 

We study the pairing symmetries for the hole-doping level $\delta$ within the range $\delta\in(0, 0.25)$, corresponding to the particle number range $n=1.5-\delta\in(1.25, 1.5)$. In this $D_4$-symmetric system, $s$-wave, $p$-wave and $d$-wave pairings are allowed. For each $\delta$, the pairing symmetry is studied for all $U<U_c$.  We find that for $\delta\leq 0.12$ the pairing symmetry is $s^{\pm}$; and for $\delta>0.12$ the pairing symmetry is $d_{xy}$, as illustrated in Fig.~\ref{rpa}(a). The obtained pairing symmetries do not rely on the value of $U$ as long as $U<U_c$. As clarified below, the variation of the pairing symmetry with the doping level is caused by the change of the FS-nesting vector.

For the case of $\delta\leq 0.12$, we present the results in Fig. \ref{rpa} (b)-(d) for a typical doping level $\delta=0.1$.  For this doping, the distribution of the spin susceptibility $\chi^{(s)}(\bm{q})$ over the BZ for $U<U_c$ is shown in Fig. \ref{rpa} (b). The distribution is just maximized at the FS-nesting vector for this doping, i.e. $\bm{Q}_1\approx(0.67\pi,0.67\pi)$, which is between the $\eta$ and $\gamma$ pockets. Fig. \ref{rpa} (c) shows the $U$-dependent $\lambda$ for the various leading pairing symmetries. The $\lambda$ for these pairing symmetries enhance promptly with the enhancement of $U$ for $U<U_c=0.7$ eV. The leading and subleading pairing symmetries are the $s$-wave and the $d_{xy}$-wave for all the $U$ parameters, respectively. The distribution of the gap function of the leading $s$-wave pairing
state on the FSs is shown in Fig. \ref{rpa} (d). This gap function changes sign between the nested patches on the $\eta$- and $\gamma$- pockets, leading to an $s^\pm$-wave pairing.

For the case of $\delta>0.12$, we present the results in Fig. \ref{rpa} (e)-(g) for a typical doping level $\delta=0.23$. For this doping, the distribution of the spin susceptibility $\chi^{(s)}(\bm{q})$ over the BZ for $U<U_c=0.86$ eV is shown in Fig. \ref{rpa} (e). The distribution is also maximized at the corresponding FS-nesting vector, $\bm{Q}_2\approx(0.9\pi,0.2\pi)$, which is between the different parts of the $\gamma$ pocket. Fig. \ref{rpa} (f) shows the $U$-dependence of $\lambda$ for different pairing symmetries, which suggests that the leading and subleading pairing symmetries are the $d_{xy}$-wave and $s$-wave, respectively. The distribution of the gap function of the leading pairing state on the FSs is shown in Fig.~\ref{rpa} (g). In Fig. \ref{rpa} (g), the gap function changes sign with every 90$^\circ$ rotation. It also changes sign upon mirror reflection about the $xz$ or the $yz$ plane, giving rise to gap nodes on the $x$- and $y$- axes. Such a gap function belongs to the $d_{xy}$-wave pairing symmetry. More results are given in Appendix B. 

\paragraph{{\color{blue} The $d_{xy}$-Wave Pairing:}} Since the ARPES measurement suggests the hole-doping level to be about $0.21$, we focus on the typical case of $\delta=0.23$ hereby. 
More details of the spin-fluctuation pattern and the pairing nature for this doping level are presented below.

\begin{figure}[htbp]
		\centering
		\includegraphics[width=0.48\textwidth]{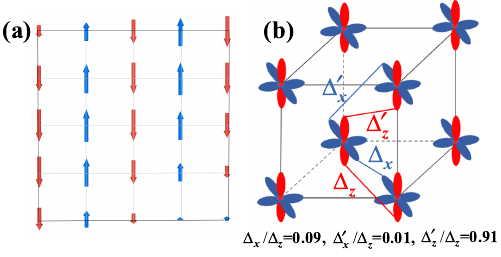}
		\caption{(color online) The results in the real space for $\delta=0.23$ and $U=0.75$ eV. (a) Schematic of the short range spin-fluctuation pattern of the $d_{z^2}$ orbital within each NiO$_2$ layer.  (b) Schematic of the real space pairing for the $d_{xy}$-wave.}
		\label{d_xy}
\end{figure}
 
In the RPA approach, the spin-fluctuation pattern is obtained from the eigenvector belonging to the largest eigenvalue of the spin susceptibility $\chi^{(s)}(\bm{Q})$. We find that in a unit cell, the dominant fluctuating magnetic moments are the intraorbital ones of the $d_{z^2}$ orbital in both layers, i.e. $m_{t/bz}$, with $m_{tz}=m_{bz}$, suggesting an interlayer ferromagnetic pattern. 
The spin-fluctuation pattern of the $d_{z^2}$ orbital over the lattice in one layer is illustrated in Fig.~\ref{d_xy} (a). In each layer, the fluctuating magnetic moment $m_{i}$ at the $i$-th site is given as $m_{i}=m_{0}\cos(\bm{Q}\cdot\bm{R}_i)$. This formula gives the stripe pattern illustrated in Fig.~{\ref{d_xy}}(a). 

In the $d_{xy}$-wave pairing state, the strongest pairing amplitude is located on the patches connected by the nesting vector $\bm{Q}_2$ within the $\gamma$ pocket, with the gap signs on these nested patches opposite to each other, as shown in Fig.~\ref{rpa} (g). Such a gap-sign structure maximizes the energy gain by superconducting condensation. Note that, different from the cases in bulk La$_3$Ni$_2$O$_7$ under HP~\cite{Wang2023LNO} and the iron-based superconductors, the FS-nesting here is not between different pockets but between different patches within one pocket, similar with the case in the cuprates. Such FS-nesting can also favor SC when the nested patches are straight enough, see Fig.~\ref{rpa} (g).

For convenience in presenting the real-space pairing configuration, we illustrate the few dominant pairing amplitudes on the few neighboring bonds in Fig.~\ref{d_xy}(b). Our RPA results suggest that all interorbital pairing components are negligible and thus all the shown pairing components are intraorbital ones. For intraorbital pairings, constrained by the $D_4$ point group symmetry, the interlayer pairing on the vertical bonds and the intralayer pairing on the nearest-neighbor bonds are forbidden by the $d_{xy}$ pairing symmetry. Our RPA results yield that the strongest pairing component is the intralayer NNN pairing for the $d_{z^2}$-orbital, i.e. the $\Delta_z$ shown in Fig.~\ref{d_xy}(b). The reason for this lies in two aspects. Firstly, as shown in 
Fig.~\ref{rpa} (g), the dominant pairing amplitude is located on the $\gamma$ pocket, which is dominated by the $d_{z^2}$ orbital. Secondly, the nearest pairing component allowed by the $d_{xy}$-wave pairing symmetry is the NNN one. The next strongest pairing component is the interlayer NNN pairing for the $d_{z^2}$-orbital, i.e. the $\Delta_z^{\prime}$ shown in Fig.~\ref{d_xy}(b), with $\Delta_z^{\prime}/\Delta_z= 0.91$. The fact that the intralayer and interlayer NNN pairing amplitudes are nearly equal originates from that the strongest pairing amplitude shown in Fig.~\ref{rpa} (g) is located on the bonding $d_{z^2}$ band in which the orbital weights from both layers are equally mixed. In addition, there are weak intralayer and interlayer NNN pairing components for the $d_{x^2-y^2}$ orbital, i.e. $\Delta_x=0.09\Delta_z$ and $\Delta_x^{\prime}=0.01\Delta_z$. All the other pairing components are negligibly weak. 

\paragraph{\color{blue}Discussion and conclusions:}
The band feature shown in Fig.~\ref{tb}(a) that the bonding-$d_{z^2}$ band top crosses the Fermi level is consistent with the ARPES. The spin-fluctuation in the stripe pattern shown in Fig.~\ref{d_xy} (a) can be detected by the soft X-ray scattering experiment, although this spin pattern might only be short-ranged. The NNN-pairing $d_{xy}$-wave SC obtained here is consistent with the fact that the $T_c$ enhances with the shrinking of the in-plane lattice constants~\cite{Ko2024signature}. This pairing symmetry can be verified in the ARPES experiment by the presence of gap nodes on the $x$- and $y$- axes. In addition, it can be identified by phase-sensitive experiments.
Now, the possibility of the $d_{xy}$-wave pairing or the $s^{\pm}$-wave pairing is not excluded by the existing ARPES results in Ref. \cite{shen2025anomalous}, where no SC gap node is observed within a small momentum range near the node of FS where the $\alpha$ and $\beta$ pockets touch each other. 

In conclusion, we have constructed a 2D bilayer Ni-$(3d_{z^2},3d_{x^2-y^2})$-orbital TB model, which well fits the DFT band structure of the La\textsubscript{3}Ni\textsubscript{2}O\textsubscript{7} ultrathin film grown on the LSAO substrate at AP. Our RPA calculations provide hole-doping $\delta$ dependent pairing symmetries. For $\delta\leq 0.12$ and $\delta> 0.12$, the leading pairing symmetries are the $s^{\pm}$- and the $d_{xy}$- respectively. This difference of the pairing symmetry originates from the different FS-nesting situations. For the hole-doping level relevant to real material, the $d_{xy}$-wave pairing is dominant. This pairing is induced by the nesting between the different patches within the $\gamma$ pocket. The real-space pairing pattern of the obtained $d_{xy}$-wave pairing is dominated by the NNN pairing of the $d_{z^2}$ orbitals. 

Finally, we have some comments. The frequency dependence has not been considered yet since it will lead to much higher computational cost. Some methods that may consider the electron correlation more sufficiently than RPA have not been considered either. These will be considered in future studies. 

~~~~~~~~~~~~~

\section*{Data availability}
All data are displayed in the main text and Appendix. 
Data within this paper are available from the corresponding author upon reasonable request. 

\section*{Acknowledgements}
We are grateful to the stimulating discussions with Jia-Heng Ji and Ze-Yu Chen. This work is supported by the National Natural Science Foundation of China (NSFC) under the Grant No. 12234016, 12074031 and 12204038.

\begin{widetext}

\appendix

\renewcommand{\theequation}{A\arabic{equation}}
\renewcommand{\thefigure}{A\arabic{figure}}
\setcounter{equation}{0}
\setcounter{figure}{0}

\section{Appendix A. The Standard RPA Approach}

In the RPA framework, the SC is driven by the spin (or charge) fluctuation whose propagator is given by the spin (or charge) susceptibility renormalized up to the RPA level. 
We first consider the bare susceptibility for the non-interacting case, defined as 
\begin{equation} \label{eq_appendix_chi0_tau}
    \chi^{(0)pq}_{st}(\bm{q},\tau) \equiv \frac{1}{N} \sum_{\bm{k}_{1},\bm{k}_2} \left\langle T_{\tau} c^{\dag}_{p}(\bm{k}_1,\tau) c_{q}(\bm{k}_1+\bm{q},\tau) c^{\dag}_{s}(\bm{k}_2+\bm{q},0) c_{t}(\bm{k}_2,0) \right\rangle_0,
\end{equation}
where $N$ is the number of unit-cells in the lattice, $p$, $q$, $s$ and $t$ are effective orbital indices. 
Transformed to the frequency space, the bare susceptibility is expressed as
\begin{equation} \label{eq_appendix_chi0_omega}
    \chi^{(0)pq}_{st}(\bm{q},\mathrm{i}\omega_n) = \frac{1}{N} \sum_{\bm{k},m_{1},m_{2}} \xi^{m_1\ast}_{p}(\bm{k}) \xi^{m_2}_{q}(\bm{k}+\bm{q}) \xi^{m_2\ast}_{s}(\bm{k}+\bm{q}) \xi^{m_1}_{t}(\bm{k}) \frac{n_F(\varepsilon^{m_2}_{\bm{k}+\bm{q}})-n_F(\varepsilon^{m_1}_{\bm{k}})}{\mathrm{i}\omega_n+\varepsilon^{m_1}_{\bm{k}}-\varepsilon^{m_2}_{\bm{k}+\bm{q}}},
\end{equation}
where $m_{1,2}$ are band indices, $\varepsilon^{b}_{\bm{k}}$ and $\xi^{b}_{p}(\bm{k})$ are the $m$-th eigenvalue and the $p$-th value in the $m$-th eigenvector of $H_{\bm{k}}$, respectively, and $n_F$ is the Fermi-Dirac distribution function. 

Then we consider the spin and charge susceptibilities for the interacting case, defined as 
\begin{align}
    & \chi^{(c)pq}_{st}(\bm{q},\tau) \equiv \frac{1}{2N} \sum_{\bm{k}_1,\bm{k}_2,\sigma_1,\sigma_2} \left\langle T_{\tau} c^{\dag}_{p\sigma_1}(\bm{k}_1,\tau) c_{q\sigma_1}(\bm{k}_1+\bm{q},\tau) c^{\dag}_{s\sigma_2}(\bm{k}_2+\bm{q},0) c_{t\sigma_2}(\bm{k}_2,0) \right\rangle, \label{eq_appendix_chic_tau} \\
    & \chi^{(s^z)pq}_{st}(\bm{q},\tau) \equiv \frac{1}{2N} \sum_{\bm{k}_1,\bm{k}_2,\sigma_1,\sigma_2} \sigma_1\sigma_2 \left\langle T_{\tau} c^{\dag}_{p\sigma_1}(\bm{k}_1,\tau) c_{q\sigma_1}(\bm{k}_1+\bm{q},\tau) c^{\dag}_{s\sigma_2}(\bm{k}_2+\bm{q},0) c_{t\sigma_2}(\bm{k}_2,0) \right\rangle, \label{eq_appendix_chisz_tau} \\
    & \chi^{(s^{+-})pq}_{st}(\bm{q},\tau) \equiv \frac{1}{N} \sum_{\bm{k}_1,\bm{k}_2} \left\langle T_{\tau} c^{\dag}_{p\uparrow}(\bm{k}_1,\tau) c_{q\downarrow}(\bm{k}_1+\bm{q},\tau) c^{\dag}_{s\downarrow}(\bm{k}_2+\bm{q},0) c_{t\uparrow}(\bm{k}_2,0) \right\rangle, \label{eq_appendix_chispm_tau} \\
    & \chi^{(s^{-+})pq}_{st}(\bm{q},\tau) \equiv \frac{1}{N} \sum_{\bm{k}_1,\bm{k}_2} \left\langle T_{\tau} c^{\dag}_{p\downarrow}(\bm{k}_1,\tau) c_{q\uparrow}(\bm{k}_1+\bm{q},\tau) c^{\dag}_{s\uparrow}(\bm{k}_2+\bm{q},0) c_{t\downarrow}(\bm{k}_2,0) \right\rangle, \label{eq_appendix_chismp_tau}
\end{align}
where the superscript $(c)$ corresponds to the charge susceptibility and $(s^z)$, $(s^{+-})$ and $(s^{-+})$ correspond to the different channels of the spin susceptibility. For a state with spin SU(2) symmetry, $\chi^{(s^z)}=\chi^{(s^{+-})}=\chi^{(s^{-+})}\equiv\chi^{(s)}$. 
We consider the multi-orbital Hubbard Hamiltonian 
\begin{equation} \label{eq_appendix_Hubbard}
    \begin{aligned}
       H = & H_{\text{TB}}+U \sum_{i\mu\alpha} n_{i\mu\alpha\uparrow}n_{i\mu\alpha\downarrow} + V \sum_{i\mu} n_{i\mu{}z}n_{i\mu{}x} \\ & + J_H \sum_{i\mu} \left[ \left( \sum_{\sigma\sigma^{\prime}} c^{\dag}_{i\mu{}z\sigma}c^{\dag}_{i\mu{}x\sigma^{\prime}}c_{i\mu{}z\sigma^{\prime}}c_{i\mu{}x\sigma} \right) + \left( c^{\dag}_{i\mu{}z\uparrow}c^{\dag}_{i\mu{}z\downarrow}c_{i\mu{}x\downarrow}c_{i\mu{}x\uparrow} + \mathrm{h.c.} \right) \right].
    \end{aligned}
\end{equation}
Note that the form in Eq. (\ref{eq_appendix_Hubbard}) is equivalent to the form 
\begin{equation}
    \begin{aligned}
        H = & H_{\text{TB}}+U \sum_{i\mu\alpha} n_{i\mu\alpha\uparrow}n_{i\mu\alpha\downarrow} + \left(V-J_H\right) \sum_{i\mu\sigma} n_{i\mu{}z\sigma}n_{i\mu{}x\sigma} + V \sum_{i\mu\sigma} n_{i\mu{}z\sigma}n_{i\mu{}x\bar{\sigma}} \\ & + J_H \sum_{i\mu} \left[ - \sum_{\sigma\sigma^{\prime}} \left( S^{+}_{i\mu{}z}S^{-}_{i\mu{}x} + S^{-}_{i\mu{}z}S^{+}_{i\mu{}x} \right) + \left( c^{\dag}_{i\mu{}z\uparrow}c^{\dag}_{i\mu{}z\downarrow}c_{i\mu{}x\downarrow}c_{i\mu{}x\uparrow} + \mathrm{h.c.} \right) \right]
   \end{aligned}
\end{equation}
Then the renormalized spin and charge susceptibilities up to the RPA level can be obtained by the following Dyson's equations 
\begin{align}
    & \chi^{(s)}(\bm{q},\mathrm{i}\omega_n) = \left( I - \chi^{(0)}(\bm{q},\mathrm{i}\omega_n) U^{(s)} \right)^{-1} \chi^{(0)}(\bm{q},\mathrm{i}\omega_n), \label{eq_appendix_chis_Dyson} \\
    & \chi^{(c)}(\bm{q},\mathrm{i}\omega_n) = \left( I + \chi^{(0)}(\bm{q},\mathrm{i}\omega_n) U^{(c)} \right)^{-1} \chi^{(0)}(\bm{q},\mathrm{i}\omega_n). \label{eq_appendix_chic_Dyson}
\end{align}
Here, $\chi^{(0)}$, $\chi^{(s/c)}$ and $U^{(s/c)}$ are viewed as matrices, with the indices in the superscript combined in the row index and the indices in the subscript combined in the column index. $U^{(s)}$ and $U^{(c)}$ take the forms
\begin{align}
    & U^{(s)pq}_{st} = 
    \begin{cases}
        U, & p=q=s=t; \\
        J_H, & (p=q)\neq(s=t) \text{ and } (p,q,s,t\leq2\text{ or }p,q,s,t>2); \\
        J_H, & (p=s)\neq(q=t) \text{ and } (p,q,s,t\leq2\text{ or }p,q,s,t>2); \\
        V, & (p=t)\neq(q=s) \text{ and } (p,q,s,t\leq2\text{ or }p,q,s,t>2);
    \end{cases} \label{eq_appendix_Us} \\
    & U^{(c)pq}_{st} = 
    \begin{cases}
        U, & p=q=s=t; \\
        2V-J_H, & (p=q)\neq(s=t) \text{ and } (p,q,s,t\leq2\text{ or }p,q,s,t>2); \\
        J_H, & (p=s)\neq(q=t) \text{ and } (p,q,s,t\leq2\text{ or }p,q,s,t>2); \\
        2J_H-V, & (p=t)\neq(q=s) \text{ and } (p,q,s,t\leq2\text{ or }p,q,s,t>2);
    \end{cases} \label{eq_appendix_Uc}
\end{align}
Only $\chi^{(s/c)}(\bm{k},\mathrm{i}\omega_n=0)\equiv\chi^{(s/c)}(\bm{k})$ is considered in this study since it is most important. 
When the interaction strength $U$ is lager than the critical interaction strength $U_c$, $\chi^{(s)}$ or $\chi^{(c)}$ will diverge, forming spin or charge order. In this study, when increasing $U$ from 0, $\chi^{(s)}$ will first diverge. 
When $U<U_c$, SC is mediated by the spin fluctuation. The spin-fluctuation pattern is obtained from the eigenvector belonging to the largest eigenvalue of the spin susceptibility $\chi^{(s)}(\bm{Q})$. 

Considering the scattering process of a Cooper pair with the total momentum equals to 0, we can get the effective attraction between electrons in the RPA level, 
\begin{equation} \label{eq_appendix_VRPA_0}
    V^{\text{RPA}}_{\text{eff}} = \frac{1}{N} \sum_{\bm{k}_1,\bm{k}_2,p,q,s,t} \Gamma^{pq}_{st}(\bm{k}_1,\bm{k}_2) c^{\dag}_{p}(\bm{k}_1) c^{\dag}_{q}(-\bm{k}_1) c_{s}(-\bm{k}_2)c_{t}(\bm{k}_2),
\end{equation}
where $\Gamma^{pq}_{st}(\bm{k}_1,\bm{k}_2)$ is the effective vertex. 
We can consider the singlet pairing channel and the triplet pairing one respectively. The effective vertices are expressed as
\begin{align}
    \Gamma^{(s)pq}_{st}(\bm{k}_1,\bm{k}_2) = & \left(\frac{U^{(c)}+3U^{(s)}}{4}\right)^{pt}_{qs} + \frac{1}{4} \left( 3U^{(s)}\chi^{(s)}(\bm{k}_{1}-\bm{k}_2)U^{(s)} - U^{(c)}\chi^{(c)}(\bm{k}_{1}-\bm{k}_2)U^{(c)} \right)^{pt}_{qs} \nonumber \\ & + \frac{1}{4} \left( 3U^{(s)}\chi^{(s)}(\bm{k}_{1}+\bm{k}_2)U^{(s)} - U^{(c)}\chi^{(c)}(\bm{k}_{1}+\bm{k}_2)U^{(c)} \right)^{ps}_{qt}, \label{eq_appendix_Gamma_s} \\
    \Gamma^{(t)pq}_{st}(\bm{k}_1,\bm{k}_2) = & \left(\frac{U^{(c)}-U^{(s)}}{4}\right)^{pt}_{qs} - \frac{1}{4} \left( U^{(s)}\chi^{(s)}(\bm{k}_{1}-\bm{k}_2)U^{(s)} + U^{(c)}\chi^{(c)}(\bm{k}_{1}-\bm{k}_2)U^{(c)} \right)^{pt}_{qs} \nonumber \\ & +\frac{1}{4} \left( U^{(s)}\chi^{(s)}(\bm{k}_{1}+\bm{k}_2)U^{(s)} + U^{(c)}\chi^{(c)}(\bm{k}_{1}+\bm{k}_2)U^{(c)} \right)^{ps}_{qt}, \label{eq_appendix_Gamma_t}
\end{align}
where the superscript $(s)$ corresponds to the singlet pairing channel and $(t)$ corresponds to the triplet pairing one. 
When only the intraband pairings are considered, the effective attraction is expressed as
\begin{equation} \label{eq_appendix_Veff_st}
    V^{(s/t)}_{\text{eff}} = \sum_{\bm{k}_1,\bm{k}_2,m_1,m_2} V^{(s/t)}_{m_1m_2}(\bm{k}_1,\bm{k_2}) c^{\dag}_{m_1}(\bm{k}_1) c^{\dag}_{m_1}(-\bm{k}_1) c_{m_2}(-\bm{k}_2) c_{m_2}(\bm{k}_2),
\end{equation}
where 
\begin{equation} \label{eq_appendix_VsVt}
    V^{(s/t)}_{m_1m_2}(\bm{k}_1,\bm{k}_2) = \sum_{p,q,s,t} \Gamma^{(s/t)pq}_{st}(\bm{k}_1,\bm{k}_2) \xi^{m_1\ast}_{p}(\bm{k}_1) \xi^{m_1\ast}_{q}(-\bm{k}_1) \xi^{m_2}_{s}(-\bm{k}_2) \xi^{m_2}_{t}(\bm{k}_2).
\end{equation}
Then the gap function and the critical temperature $T_c$ can be obtained by solving the linearized gap equation. 
\begin{equation} \label{eq_appendix_gap_equation}
    \Delta^{(s/t)}_{m_1}(\bm{k}) = \frac{1}{N} \sum_{\bm{q},m_2} \frac{V^{(s/t)}_{m_1m_2}(\bm{k},\bm{q}) \tanh{\frac{\varepsilon_{m_2}(\bm{q})}{2T_c}}}{2\varepsilon_{m_2}(\bm{q})} \Delta_{m_2}(\bm{q}). 
\end{equation}
Here, the summation is done within the full BZ. 
Solving this equation is equivalent to solving the eigenvalues and eigenvectors of a matrix, where $T_c$ is treated as a variable. If and only if the variable value is the critical temperature, the maximum eigenvalue of the matrix is equal to 1 and the corresponding eigenvector is the gap function $\Delta$. 
However, the matrix in eq. (\ref{eq_appendix_gap_equation}) is not Hermitian. To make it an eigenvalue problem of a Hermitian matrix, we define $\tilde{\Delta}_{m_1}(\bm{k})\equiv\Delta_{m_1}(\bm{k})\sqrt{\frac{\tanh{\frac{\varepsilon_{m_1}(\bm{k})}{2T_c}}}{2\varepsilon_{m_1}(\bm{k})}}$, then eq. (\ref{eq_appendix_gap_equation}) is transformed to
\begin{equation} \label{eq_appendix_gap_equation_Hermitian}
    \tilde{\Delta}_{m_1}(\bm{k}) = \frac{1}{N} \sum_{\bm{q},m_2} V_{m_1m_2}(\bm{k},\bm{q}) \sqrt{\frac{\tanh{\frac{\varepsilon_{m_1}(\bm{k})}{2T_c}}}{2\varepsilon_{m_1}(\bm{k})}} \sqrt{\frac{\tanh{\frac{\varepsilon_{m_2}(\bm{q})}{2T_c}}}{2\varepsilon_{m_2}(\bm{q})}} \tilde{\Delta}_{m_2}(\bm{q}). 
\end{equation}
Now, solving $T_c$ and the gap function is equivalent to the eigenvalue problem of a Hermitian matrix, where $T_c$ is also treated as a variable. If and only if the variable value is the critical temperature, the maximum eigenvalue of the matrix is equal to 1 and the gap function $\Delta$ can be obtained from the corresponding eigenvector. 

In eq. (\ref{eq_appendix_gap_equation}), the momenta $\bm{q}$ near the FS contribute most to the summation. Thus, it is reasonable to do the summation only over the momenta $\bm{q}$ near the FS. Then eq. (\ref{eq_appendix_gap_equation}) is simplified as 
\begin{equation} \label{eq_appendix_gap_equation_FS}
    -\frac{1}{\left(2\pi\right)^2} \sum_{m_2} \oint_{\text{FS}} \mathrm{d}\bm{q}_{\parallel} \frac{V_{m_1m_2}(\bm{k},\bm{q})}{v_F^{m_2}(\bm{q})} \Delta_{\bm{q}m_2} = \lambda \Delta_{\bm{k}m_1}, 
\end{equation}
where the integration is done along the $m_2$-FS in each term of the summation, $v_F^{m_2}$ is the Fermi velocity of the $m_2$ band and $\bm{q}_{\parallel}$ is the component of $\bm{q}$ along the FS. 
Solving this equation is also equivalent to a eigenvalue problem. Here, $T_c$ of SC is determined by the largest eigenvalue $\lambda$ through $T_c\sim e^{-1/\lambda}$, and the pairing gap function $\Delta$ is determined by the corresponding eigenvector. 

\section{Appendix B. More Results}

In the main text, all of the pairing eigenvalues, related to $T_c$, and gap functions are obtained from eq. (\ref{eq_appendix_gap_equation_FS}), where the integration is done along the FS. We also solve eq. (\ref{eq_appendix_gap_equation_Hermitian}), where the integration is done within the full BZ, for $\delta=0.23$ and various $U$ values to obtain the $T_c$ and the gap functions of some leading pairing symmetries. 

$k_BT_c$ of some leading pairing symmetries as function of $U$ at $\delta=0.23$ is shown in Fig. \ref{allBZ} (a). The $T_c$ of the leading pairing symmetries increases as $U$ increases. Also, the $T_c$ of $d_{xy}$-wave pairing is much larger than the one of $d_{x^2-y^2}$-wave or $s$-wave for all $U$ calculated. 

The distribution of the gap function with the highest $T_c$ for $U=0.75\text{ eV}$ and $\delta=0.23$ is shown in Fig. \ref{allBZ} (b). For each momentum $\bm{k}$, the plotted value of the gap function is the one of the $m_{\bm{k}}$-th band $\Delta_{\bm{k}m_{\bm{k}}}$, whose absolute value is the largest among the values of all bands. Here, the $m_{\bm{k}}$ for different $\bm{k}$ can be different. 
This gap function displays $d_{xy}$ symmetry, where the strongest pairing amplitude is located near the $\gamma$ pocket while the pairing amplitudes of other momenta are relatively weak. 

All of the above results are qualitatively consistent with the ones in the main text. 

\begin{figure}[htbp]
		\centering
		\includegraphics[width=0.8\textwidth]{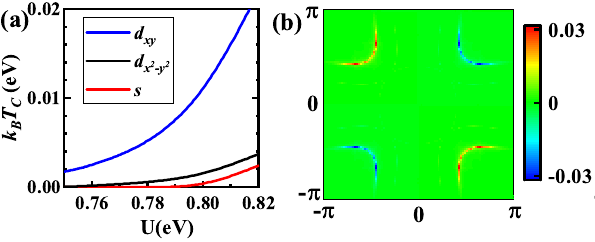}
		\caption{(color online) (a) The $k_BT_c$ of different pairing symmetries as function of $U$ at $\delta=0.23$. (b) The distribution of the $d_{xy}$-wave pairing gap function in the BZ for $\delta=0.23$ and $U=0.75\text{ eV}$.}
		\label{allBZ}
\end{figure}

\end{widetext}

\bibliography{reference}

\end{document}